\documentclass{ws-procs9x6}

\begin{document}

\title{Dark Matter Search with Direction Sensitive Scintillators}

\author{H.~SEKIYA}

\address{Department of Physics, School of Science, Kyoto University\\
Kitashirakawa, Sakyo, Kyoto 606-8502, Japan \\
E-mail: sekiya@cr.scphys.kyoto-u.ac.jp}

\author{M.~MINOWA, Y.~SHIMIZU, W.~SUGANUMA}

\address{Department of Physics, School of Science, University of Tokyo\\
7-3-1 Hongo Bunkyo-ku, Tokyo, 113-0033}

\author{Y.~INOUE}

\address{International Center for Elementary Particle Physics,
University of Tokyo\\
7-3-1 Hongo Bunkyo-ku, Tokyo, 113-0033}

\maketitle

\abstracts{
We have carried out the dark matter search with a 116g direction-sensitive
stilbene crystal in Kamioka Observatory.
With the crystal fixed to the earth, we searched the modulation of the
light output.
No modulation signal was found due to the small size of the detector crystal and the higher
background rate yet to be eliminated. 
However, it demonstrated the effectiveness of the method of
direction sensitive search for the dark matter with
an implementation of the anisotropic organic scintillation crystal. 
}
 
\section{Introduction}
The most convincing signature of the WIMPs appears in 
the direction of nuclear recoils induced by WIMPs\cite{spergel}.
Although studies on detecting the signature by measuring the
recoil directions have been carried out ever since it was
indicated to be a reliable method, 
no dark matter search had been conducted with direction
sensitive detector because of its difficulties.  
Recoil energy is only a few tens of keV and the track length
of recoil nucleus should be short.
Consequently, low pressure TPC have been studied as the 
directional WIMP detector principally, such as DRIFT and NEWAGE\cite{drinew}.
However, realistic experiments with gaseous detectors
are very challenging because they are required the
large fiducial volume and the great stability with
very fine resolution.
Therefore, it is significant to explore alternative experimental 
approach. 

It is known that scintillation efficiency of organic crystals
to heavy particles depends on the direction of the particles 
with respect to the crystallographic axes. This property makes it
 possible to propose a WIMP detector sensitive to the recoil 
direction of the nucleus\cite{org}.

We measured the carbon recoils in 
a stilbene crystal for recoil energies of 30 keV to 1 MeV and
 shown that the scintillation efficiency does vary by 7$\%$ depending 
on the direction of the recoil carbon with respect to $c'$ axis\cite{sekiya}.

\begin{figure}[t]
  \centerline{%
    \includegraphics[scale=.3]{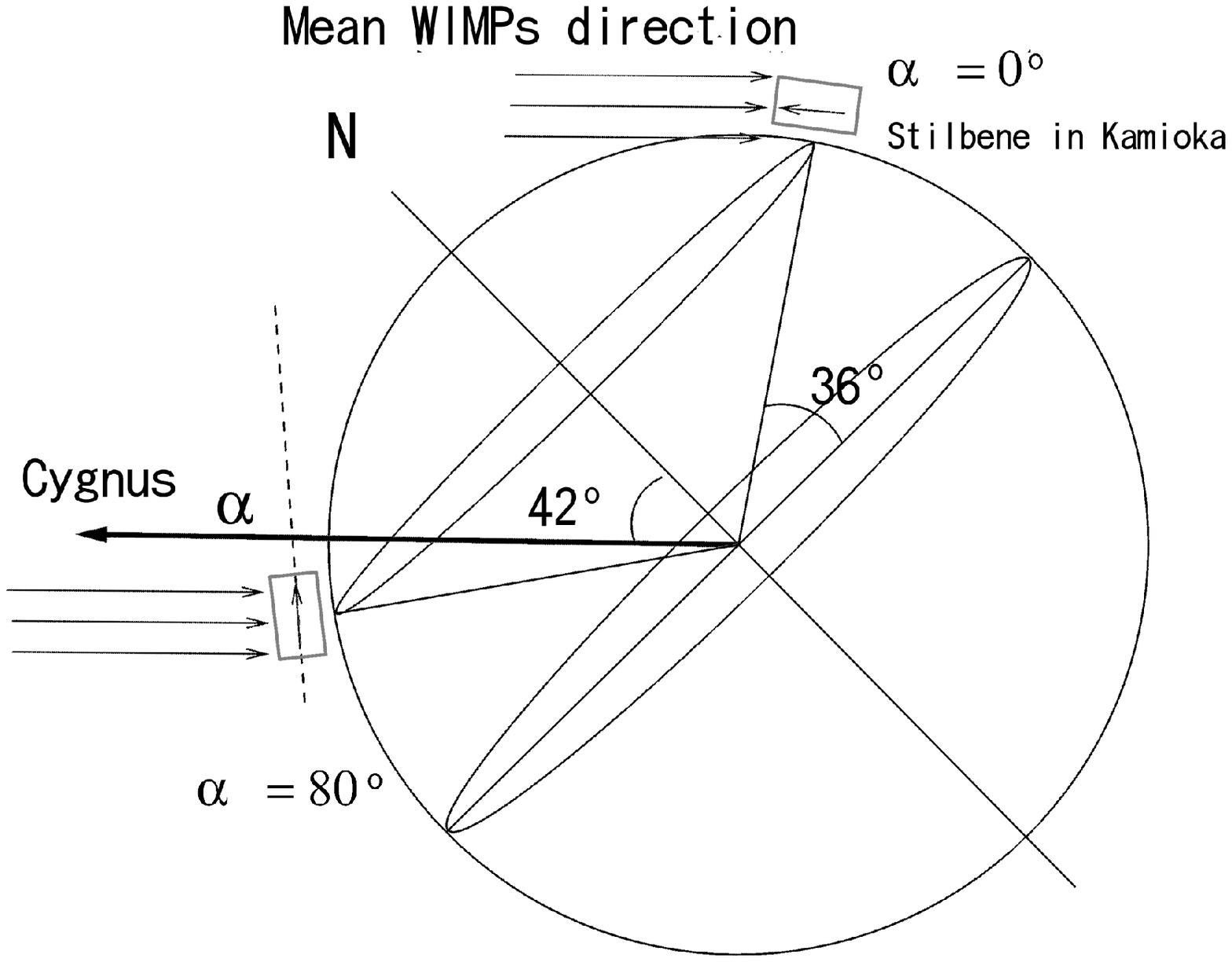}
    \hskip 1pc
    \includegraphics[scale=.5]{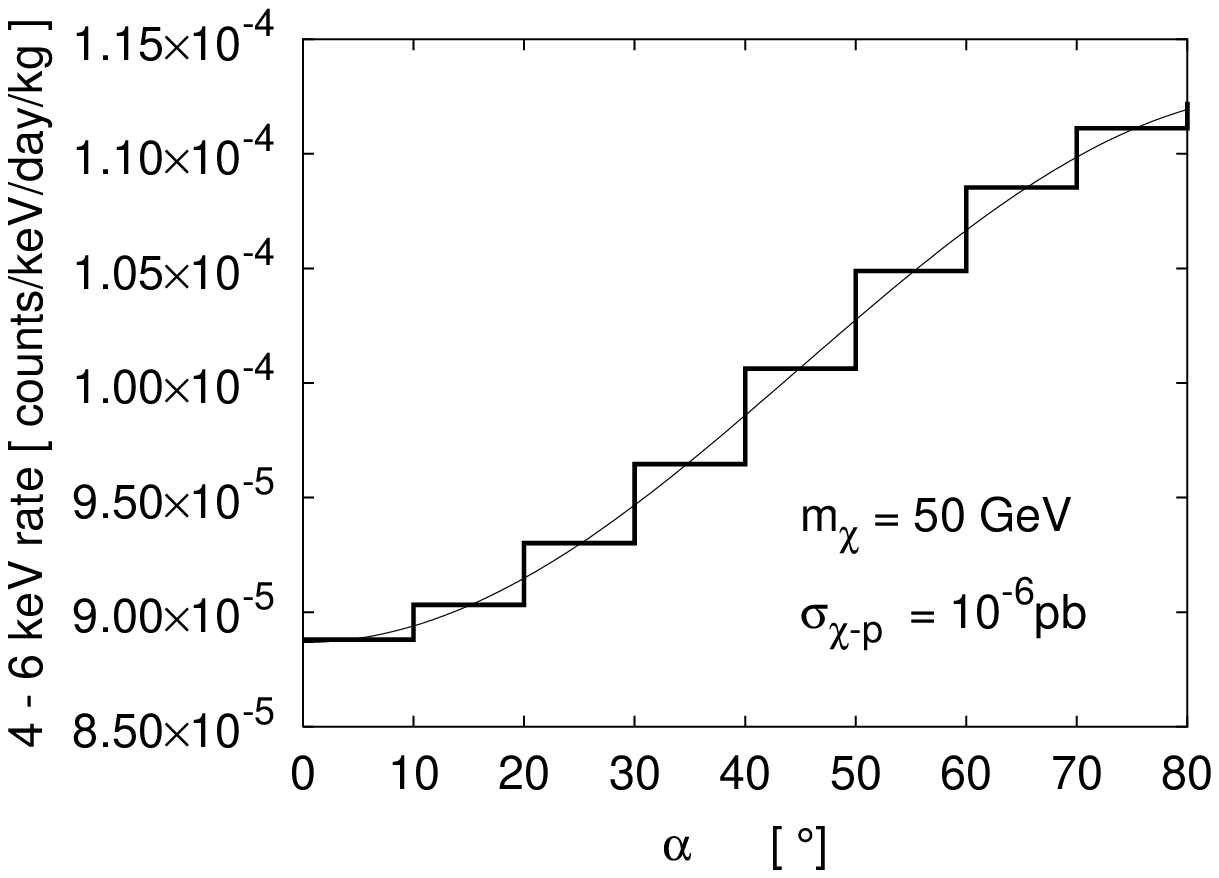}}
  \caption{Left: Schematic drawing of the experimental approach
 mentioned in the text. Right: The expected event rate for 4-6 keV
 as a function of $\alpha$. The thin line shows
 $A-S\cos2\alpha$, where $S=1.20\times10^{-5}$ counts/keV/day/kg
   and $A=1.01\times10^{-4}$ counts/keV/day/kg in this case.
The  parameters that we used in the calculation are 
 $\rho_0=0.3$ GeV/cm$^3$, $v_0=220$ km/sec, 
 WIMP-proton spin independent cross section $\sigma_{\chi-p}=10^{-6}$ pb,
 and $m_{\chi}=50$ GeV.}
  \label{fig:stilrotate}
\end{figure}

Then, we estimated the response to WIMPs when stilbene crystals installed
in Kamioka\cite{noon}. As illustrated in Fig.\ \ref{fig:stilrotate}, 
a suitable arrangement for the stilbene crystal is to fix 
the detector with the $c'$ axis in parallel to the horizontal plane 
and towards the North assuming the WIMP halo is an 
isothermal sphere. In that case, the mean incident angle of the WIMP with respect 
to $c'$ axis, $\alpha$,  varies about $80^\circ$ within a sidereal daily period.

The expected event rate for 4-6 keV region as a function of
$\alpha$ calculated by Monte Carlo method is shown in Fig.\ \ref{fig:stilrotate}.
As indicated, the variation can be well fitted by a function $A-S\cos2\alpha$. 

As the next step, we have performed a pilot experiment at Kamioka to
prove the feasibility of this method. 
In this paper, we report on the measurement and its results. 

\section{The Experimental Setup}
\begin{figure}[t]
\begin{center}
    \includegraphics[scale=.4]{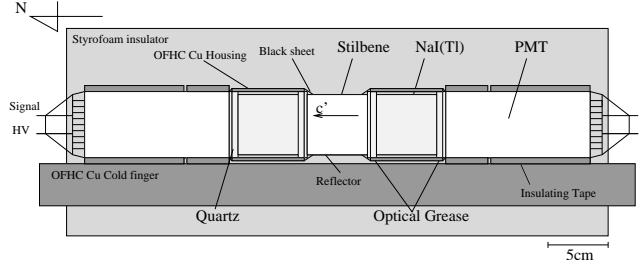}
\end{center}
  \caption{Schematic view of the detector setup.}
  \label{fig:pilot}
\end{figure}

The schematic view of the detector assembly is shown
in Fig.\ \ref{fig:pilot}.
The $\phi50{\rm\,mm}\times50{\rm\,mm}$ (116\,g)
cylindrical stilbene crystal is viewed by two
Hamamatsu R8778MOD low background PMTs through two Horiba 
low background NaI(Tl) active shields.
Self coincidence of two PMTs are required 
and both PMTs are cooled down at about $-7^\circ\rm\!C$ to 
reduce dark current further.

The detector assembly is shielded with 10cm OFHC copper, 15 cm Lead, and 
20cm polyethylene. The EVOH sheets are formed into
 air tight bags filled with nitrogen gas for purging the radon gas.

The whole setup is laid with the $c'$ axis of the stilbene crystal
parallel to the north-south direction.

\section{Measurement Results}
\begin{figure}[t]
\begin{center}
    \includegraphics[scale=.5]{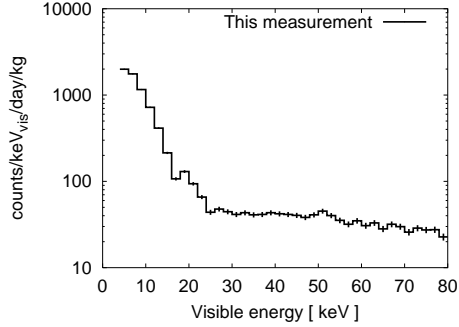}
  \caption{Low energy spectrum obtained with the stilbene crystal.}
  \label{fig:rate46}
\end{center}
\end{figure}

With the detector system, we started the measurement in October 25, 2003
and it was halted in December 11, 2003\cite{noon}.
The obtained energy spectrum with the stilbene is shown in Fig.\ \ref{fig:rate46}.

\begin{figure}[tb]
\begin{center}
    \includegraphics[scale=.8]{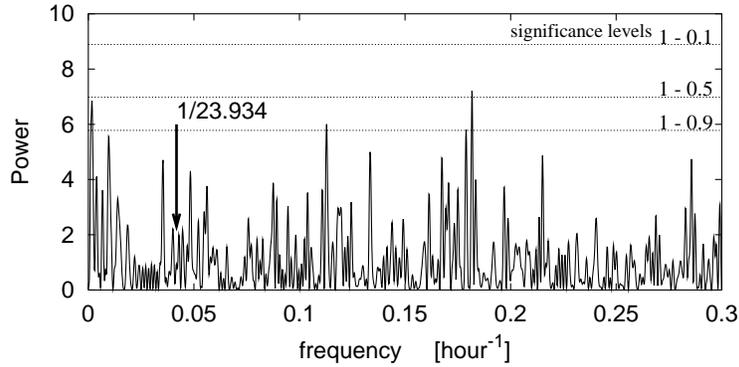}
\end{center}
  \caption{Power spectrum of the time data of event rate for 4-6 keV in
 the stilbene.}
 \label{fig:pspe}
\end{figure}

The background event rate is as high as 2000 counts/keV/day/kg for
4-6 keV region, however,
the event rate of WIMPs should change in a cycle of one sidereal day
---i.e. 23.934 hours--- independent of halo models.
Therefore, in order to search the modulation signal in frequency domain,
we derived the power spectrum of the time data of 
the event rate for 4-6 keV region during the measurements\cite{noon}
using Lomb's periodgram method\cite{ram}.
Fig.\ \ref{fig:pspe} shows the power spectrum for the frequency 
interval 0 - 0.3 hour$^{-1}$.

The signal with the frequency of $1/23.934$ hours cannot be discerned
from Fig.\ \ref{fig:pspe}.
That means the isotropic ``white noise'' events dominates the 
high rates of background events which should be eliminated.

\begin{figure}[tb]
\begin{center}
    \includegraphics[scale=.5]{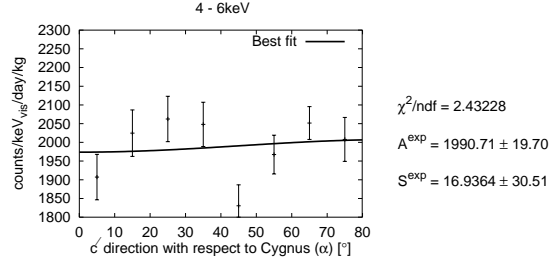}
\end{center}
  \caption{The measured event rate for 4-6 keV as a
 function of $\alpha$. Expected $A-S\cos2\alpha$ is fitted.}
 \label{fig:analF}
\end{figure}

\begin{figure}[tb]
\begin{center}
    \includegraphics[scale=.5]{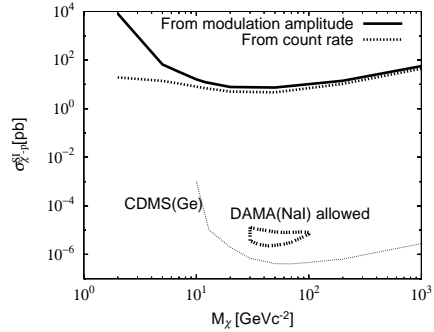}
\end{center}
  \caption{The obtained $\sigma_{\chi-p}$ limits as a function of
 WIMP mass $M_{\chi}$.}
 \label{fig:limit}
\end{figure}

For all that, we derived limits on cross sections from 
this isotropic results. Fig.\ \ref{fig:analF} shows the measured event 
rate for 4-6 keV as a function of $\alpha$. 
As indicated in Fig.\ \ref{fig:stilrotate}, the event rate should
vary as $A-S\cos2\alpha$, and both $A$ and $S$ are in proportion to 
$\sigma_{\chi-p}$. Accordingly,
from the measured unmodulated part, $A$, conventional limits 
can be derived, and from the measured modulation amplitude, $S$,
limits of direction sensitivity can be derived.
The obtained limits on $\sigma_{\chi-p}$ are shown in Fig.\ \ref{fig:limit}.
The limit is far looser than the contemporary non-directional limits,
however, it is the limit from the directional signature of WIMPs.
In addition, as the directional limit is derived from the ``signal of
WIMPs'', the better limits will be obtained with the higher statistics.

\begin{figure}[tb]
\begin{center}
    \includegraphics[scale=.5]{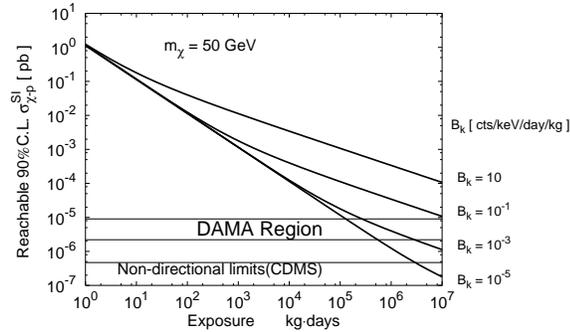}
\end{center}
  \caption{Background rate dependence of the expected sensitivity.}
 \label{fig:reablebg}
\end{figure}

Fig.\ \ref{fig:reablebg} indicates the background and exposure dependence of 
the achievable limits on $\sigma_{\chi-p}$. We see from 
Fig.\ \ref{fig:reablebg} that the directional limit will be comparable to
contemporary non-directional limit if we achieved the background
level as low as $10^{-3}$ counts/keV/day/kg. 
Stilbene crystals could potentially detect the robuster WIMP signal
arising from the earth's rotation in our galaxy as 
compared with the annual modulation signal.
 
\section{Discussions and Prospects}
\begin{figure}[tb]
\begin{center}
    \includegraphics[scale=.41]{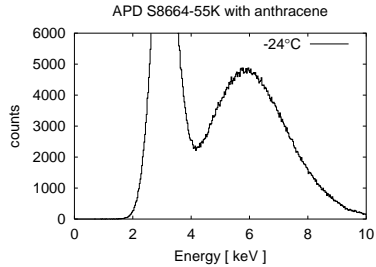}
\end{center}
  \caption{An example of the performance of the HAMAMATSU S8664-55K. 5.9 keV X-ray spectrum 
measured at $-24^{\circ}$C with anthracene crystal($15\times 15\times 15$ mm$^{3}$).}
 \label{fig:fespe}
\end{figure}

It is obvious that rather high background rate due to the radioactivity
in PMTs limits the sensitivity. The small light yield of the stilbene
is another essential problem.
Therefore, in order to overcome the difficulties, 
highly radio-pure, high quantum efficiency, high gain photon detector
is indispensable. In that respect, we focused on Avalanche
photodiodes(APD). Fig.\ \ref{fig:fespe} shows an example of the 
performance of an APD (HAMAMATSU S8664-55K) with anthracene crystal. 
5.9keV X-rays from $^{55}$Fe are clearly resolved. This result suggests
that APDs will be promising devices for organic scintillators
and will pull up the potential of the direction sensitivity.

\end{document}